\documentclass[12pt]{iopart}
\usepackage{graphicx}
\begin{document}
\title{Examining Gravitational Collapse With Test Scalar Fields}
\author{Ryo Saotome}
\address{Michigan Center for Theoretical Physics, Randall Laboratory of Physics, University of Michigan, Ann Arbor, MI 48109-1120, USA}
\ead{rsaotome@umich.edu}
\author{Ratindranath Akhoury}
\address{Michigan Center for Theoretical Physics, Randall Laboratory of Physics, University of Michigan, Ann Arbor, MI 48109-1120, USA}
\ead{akhoury@umich.edu}
\author{David Garfinkle}
\address{Dept. of Physics, Oakland University,
Rochester, MI 48309, USA}
\address{and Michigan Center for Theoretical Physics, Randall Laboratory of Physics, University of Michigan, Ann Arbor, MI 48109-1120, USA}
\ead{garfinkl@oakland.edu}
\begin{abstract}
Numerical simulations are performed of a test scalar field in a spacetime undergoing gravitational collapse.  The behavior of the scalar field near the singularity is examined and implications for generic singularities are discussed. In particular, our example is the first confirmation of the BKL conjecture for an asymptotically flat spacetime.
\end{abstract}
\maketitle
\section{Introduction}
\qquad
Gravitational collapse results in black holes which contain singularities.  The collapse process is 
not described by an analytical solution of the field equations, and the gravitational field is so strong that perturbation theory is not helpful either.  So numerical simulation is needed.  But even here, the full process of gravitational collapse is not usually examined: some simulations examine the formation of black holes but not singularities \cite{matt} 
while others examine singularities but not black holes \cite{beverly,me}.  
Note, however, that there has been some recent progress in numerical simulations capable of examining
the formation of both black holes and singularities.\cite{racz}
In the case where black holes are examined, the simulation sometimes ends when the black hole forms \cite{matt} or the simulation goes on, but the interior of the black hole is excised \cite{matt2}.
In the case where the singularity is examined, one usually imposes periodic boundary conditions, which
means that what is studied is singularity formation in a closed cosmology.  

There is a longstanding conjecture due to Belinskii, Khalatnikov, and Lifschitz (BKL) as to the general behavior of spacetime singularities.\cite{bkl}
Essentially the BKL conjecture is that as a generic singularity is approached in a comoving coordinate system, the field equations will be dominated by terms containing derivatives with respect to time, and all other terms can be neglected.  The BKL conjecture has been verified by numerical 
simulations\cite{beverly,me} for the case of singularity formation in a closed cosmology.  Since BKL
dynamics are local, it seems likely that the BKL conjecture is also true in the case of singularity formation inside a black hole (at least for part of the singularity, see \cite{eric} for a conjecture
about another part of the singularity).  However, one would like to verify the BKL conjecture by a treatment of the singularities inside black holes.  

One of the earliest studies of gravitational collapse, that of Oppenheimer and Snyder \cite{OS} was capable of also examining the singularity.  In \cite{OS} gravitational collapse is modeled as a spherically symmetric spacetime containing two parts: the Schwarzschild metric on the outside and a Friedmann-Robertson-Walker (FRW) spacetime with non-relativistic matter (dust) on the inside.  The Schwarzschild and FRW spacetimes have two different coordinate systems which must be matched along the boundary.  This difficulty can be overcome \cite{bj} through a coordinate transformation which allows the full spacetime to be covered by a single coordinate system.  However a more serious difficulty is that pressureless dust is not a good model of how matter behaves in the extreme conditions of singularity formation.  So the properties of dust singularities tell us very little about the properties of generic singularities.  

Ideally, one would like to check the BKL conjecture by treating the gravitational collapse of a more realistic matter model, like a scalar field, using a single coordinate system that follows the formation of the black hole and the singularity.  As a preliminary step in that direction, we perform numerical simulations of the behavior of a test massive scalar field on the spacetime of \cite{OS} using the coordinates of 
\cite{bj}.  The relevant equations and numerical methods are described in section 2, while results are presented and discussed in section 3.

\section{Methods}

A spherically symmetric spacetime can be put in the form \cite{P,G}
\begin{equation}
d{s^2} = -(1-{\psi ^2}) d{t^2} + 2 \psi dt dr + {r^2}(d {\theta ^2} + {\sin ^2} \theta d 
{\varphi^2})
\label{pgmetric}
\end{equation}
Here, $r$ is the usual area coordinate, while $t$ is a time coordinate chosen so that 
${\nabla ^a}t{\nabla _a}t = -1$.  We now show how to put the Schwarzschild and FRW metrics in this form.  The Schwarzschild metric in the usual coordinates is
\begin{equation}
d {s^2} = - \left ( 1 - {\frac {2 M} r}\right ) d {{\tilde t}^2} + {{\left ( 1 - {\frac {2 M} r}\right ) }^{-1}} d {r^2} + {r^2}(d {\theta ^2} + {\sin ^2} \theta d 
{\varphi^2})
\label{sch}
\end{equation}  
We wish to express $\tilde t$ as a function of $t$ and $r$ so that the metric takes the form of 
eqn. (\ref{pgmetric}).  Some straightforward but tedious algebra shows that the expression
\begin{equation}
{\tilde t} = t - 2 \left [ {\sqrt {2Mr}} + M \ln \left ( {\frac {{\sqrt r} - {\sqrt {2M}}}
{{\sqrt r} + {\sqrt {2M}}}} \right ) \right ]
\end{equation}
yields a metric of the form of eqn. (\ref{pgmetric}) with 
\begin{equation}
\psi = {\sqrt {\frac {2M} r}}
\label{schpsi}
\end{equation}
The spatially flat dust FRW spacetime has the metric
\begin{equation}
d {s^2} = - d {{\tilde t}^2} + {a^2}({\tilde t}) ( d {{\tilde r}^2} + {{\tilde r}^2} [
d {\theta ^2} + {\sin ^2} \theta d 
{\varphi^2}])
\label{frw}
\end{equation}
with the scale factor $a$ given by 
\begin{equation}
a = K {{\tilde t}^{2/3}}
\label{adust}
\end{equation}
where $K$ is a constant.  We wish to perform a coordinate transformation that gives this metric the form of eqn. (\ref{pgmetric}).  Comparing the angular parts of the metric, it is clear that we
must have $r = a {\tilde r}$.  It is then straightforward to show that with ${\tilde t} = t$ the
metric of eqn. (\ref{frw}) takes the form of eqn. (\ref{pgmetric}) with 
\begin{equation}
\psi = - {\frac {2r} {3t}}
\label{frwpsi}
\end{equation}

The collapse takes place at negative times and is complete at $t=0$.  The boundary between the
interior FRW part of the spacetime and the exterior Schwarzschild part is given by $r = {r_b}(t)$.
for some function ${r_b}(t)$.  Since the metric, and therefore $\psi$ must be continuous across
the boundary, we have 
\begin{equation}
{r_b} = {{\left ( {\textstyle {\frac 9 2}} M {t^2} \right ) }^{1/3}}
\end{equation}
Thus the spacetime metric is given by eqn. (\ref{pgmetric}) with $\psi $ given by eqn (\ref{schpsi})
for $r \ge {r_b}$ and $\psi $ given by eqn. (\ref{frwpsi}) for $r \le {r_b}$.  

We wish to simulate the Klein-Gordon equation on this spacetime assuming spherical symmetry for
the scalar field.  That is, we want a scalar field $\phi (r,t)$ that satisfies
\begin{equation}
{\nabla _a}{\nabla ^a} \phi - {m^2} \phi = 0
\label{kleingordon}
\end{equation}
For a metric of the form of eqn. (\ref{pgmetric}) the Klein-Gordon equation for a spherically symmetric scalar field becomes
\begin{eqnarray}
{\partial _t}{\partial _t} \phi &=& 2 \psi {\partial _r} {\partial _t} \phi + ({\partial _t}\psi )
({\partial _r} \phi ) + ({\partial _r} \psi ) ({\partial _t} \phi ) + {\frac {2\psi } r}
{\partial _t} \phi 
\nonumber
\\
&+& (1 - {\psi ^2}) {\frac 1 {r^2}}{\partial _r} ( {r^2} {\partial _r} \phi ) 
- 2 \psi ({\partial _r} \psi ) {\partial _r} \phi - {m^2} \phi 
\label{kg}
\end{eqnarray}
For the numerical method used, we put this equation in first order form: we introduce the quantities 
$P$ and $S$ given by 
\begin{eqnarray}
P \equiv {\partial _t} \phi
\\
S \equiv {\partial _r} \phi
\end{eqnarray}
From the definitions of $P$ and $S$ it immediately follows that
\begin{eqnarray}
{\partial _t} \phi = P
\label{dtphi}
\\
{\partial _t} S = {\partial _r} P
\label{dts}
\end{eqnarray}
while eqn. (\ref{kg}) becomes
\begin{eqnarray}
{\partial _t} P &=& 2 \psi {\partial _r} P + ({\partial _t}\psi ) S
+ ({\partial _r} \psi ) P + {\frac {2\psi } r} P
\nonumber
\\
&+& (1 - {\psi ^2}) {\frac 1 {r^2}}{\partial _r} ( {r^2} S ) 
- 2 \psi ({\partial _r} \psi ) S - {m^2} \phi 
\label{kg1}
\end{eqnarray}

The equations to be evolved are eqns. (\ref{dtphi}-\ref{kg1}).  We evolve using finite differences, 
where each scalar $f(r)$ is represented by the quantities ${f_i}=f(i\Delta r)$.  Spatial derivatives
are evaluated using standard centered differences, that is
\begin{equation}
{\partial _r}f \to {\frac {{f_{i+1}}-{f_{i-1}}} {2 \Delta r}}
\end{equation}
However, to maintain accuracy near the origin, the quantity ${r^{-2}} {\partial _r} ( {r^2} S )$
requires special treatment.  We define $S_+$ and $S_-$ by ${S_+}=({S_{i+1}}+{S_i})/2$ and 
${S_-}=({S_i}+{S_{i-1}})/2$. Correspondingly we define ${r_+} = {r_i}+(\Delta r/2)$ and 
${r_-} ={r_i} - (\Delta r/2)$.  Then we use the finite difference
\begin{equation}
{\frac 1 {r^2}}{\partial _r} ( {r^2} S ) \to {\frac 3 {{r_+ ^3} - {r_- ^3}}} \left (
{r_+ ^2} {S_+} - {r_- ^2} {S_-} \right )
\end{equation}
The quantity $\psi$ and its derivatives are evaluated analytically using the formulas in eqns. 
(\ref{schpsi}) and (\ref{frwpsi}).
Eqns. (\ref{dtphi}-\ref{kg1}) are evolved using a three step iterated Crank-Nicholson method \cite{icn}
with Kreiss-Oliger dissipation \cite{ko}.

At the origin, (gridpoint $i=1$) smoothness requires that $S$ vanish and that $\phi$ and $P$ have 
vanishing derivative with respect to $r$.  We implement this condition as
\begin{eqnarray}
{\phi _1} = (4 {\phi _2} - {\phi _3})/3
\\
{S_1}=0
\\
{P_1} = (4 {P_2}-{P_3})/3
\end{eqnarray}
At the outer boundary, we impose the condition that $\phi , \, S$ and $P$ all vanish.

\section{Results} 

\begin{figure}
\includegraphics{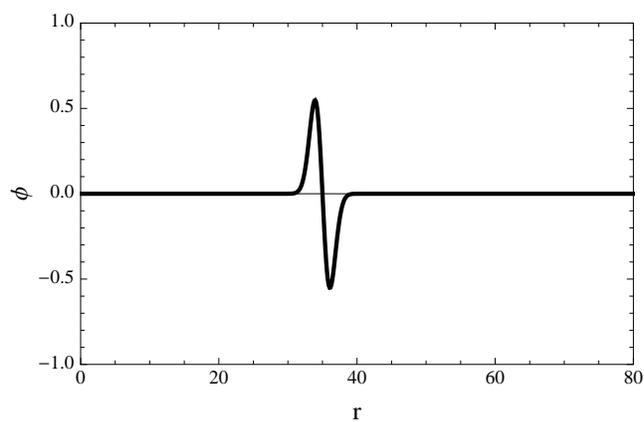}
\caption{\label{fig1}$\phi$ {\it vs} $r$ at the initial time ${t_0} = - 20$ }
\end{figure}

\begin{figure}
\includegraphics{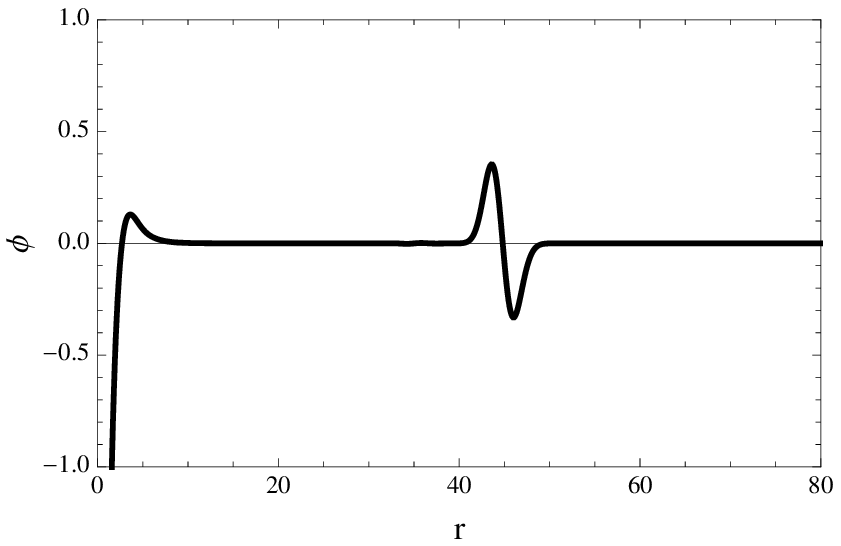}
\caption{\label{fig2}$\phi$ {\it vs} $r$ for $m=0$ at $t=-0.23$ }
\end{figure}

\begin{figure}
\includegraphics{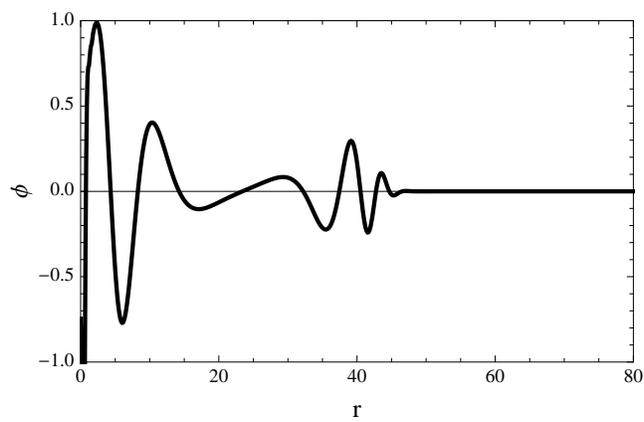}
\caption{\label{fig3}$\phi$ {\it vs} $r$ for $m=1$ at $t=-0.23$ }
\end{figure}

\begin{figure}
\includegraphics{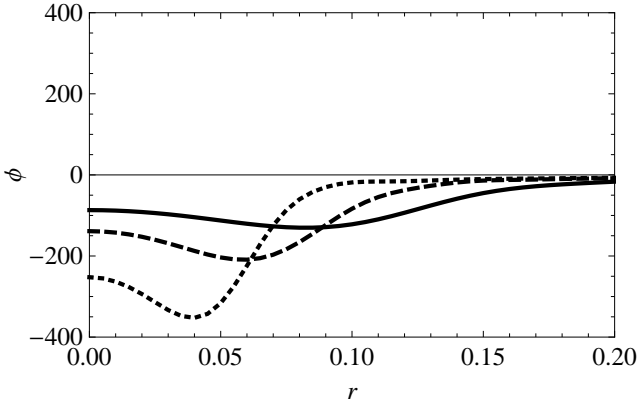}
\caption{\label{fig4}$\phi$ {\it vs} $r$ for $m=0$ at $t=-0.02$ (solid)$, \, -0.01$ (dashed) and $-0.005$ (dotted)}
\end{figure}

\begin{figure}
\includegraphics{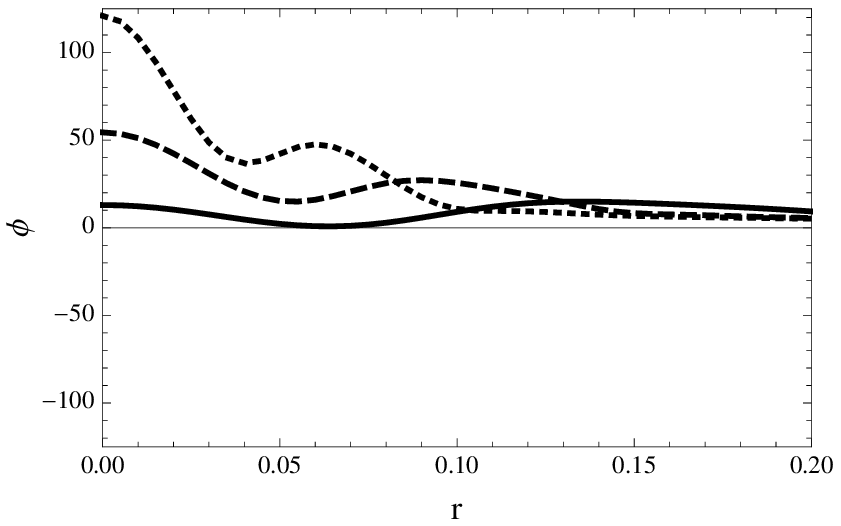}
\caption{\label{fig5}$\phi$ {\it vs} $r$ for $m=1$ at $t=-0.02$ (solid)$, \, -0.01$ (dashed) and $-0.005$ (dotted)}
\end{figure}

\begin{figure}
\includegraphics{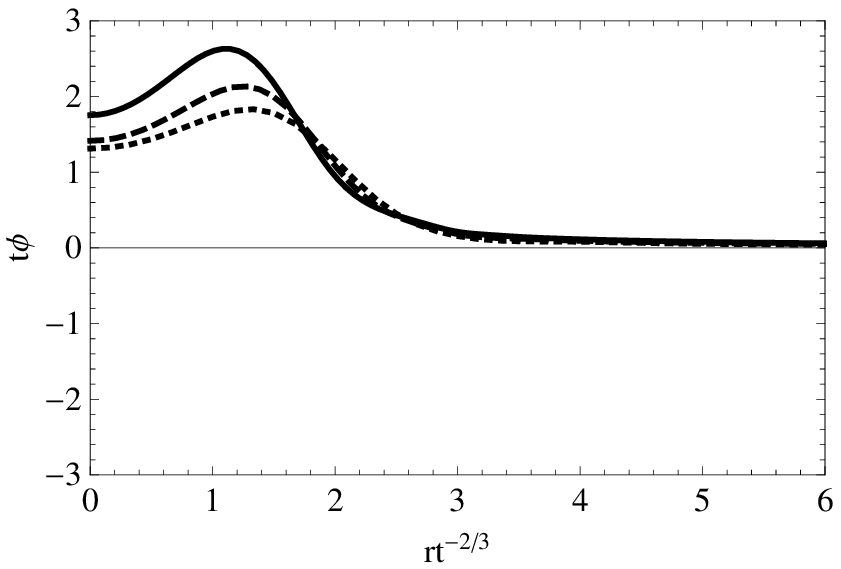}
\caption{\label{fig6}$t\phi$ {\it vs} ${t^{-2/3}}r$ for $m=0$ at $t=-0.02$ (solid)$, \, -0.01$ (dashed) and $-0.005$ (dotted)}
\end{figure}

\begin{figure}
\includegraphics{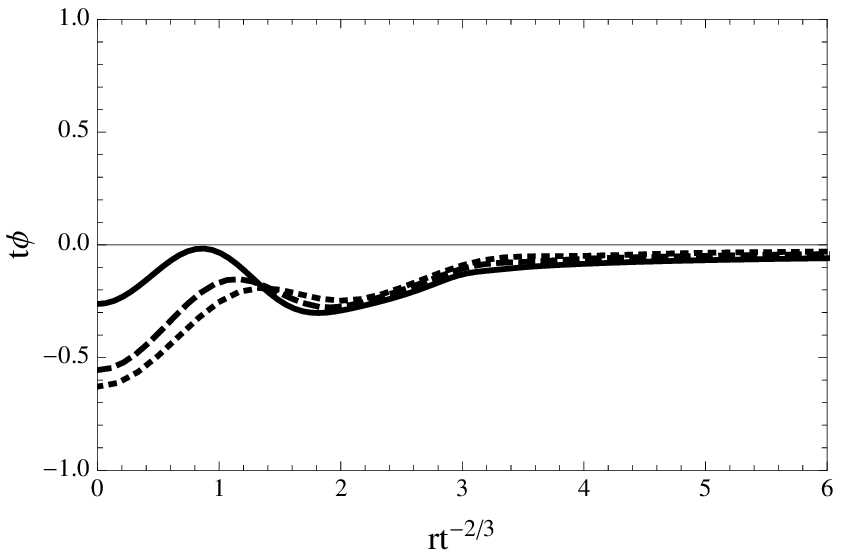}
\caption{\label{fig7}$t\phi$ {\it vs} ${t^{-2/3}}r$ for $m=1$ at $t=-0.02$ (solid)$, \, -0.01$ (dashed) and $-0.005$ (dotted)}
\end{figure}

\begin{figure}
\includegraphics{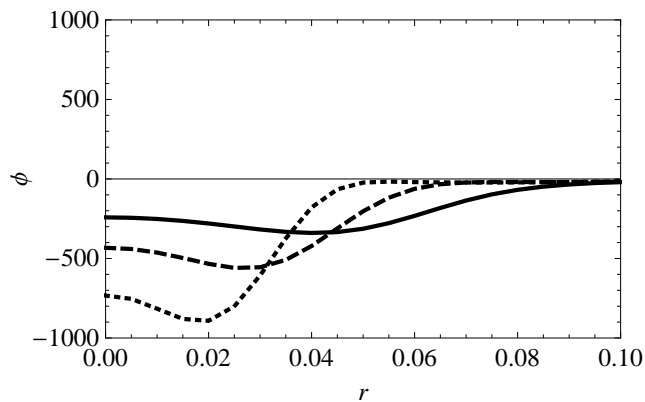}
\caption{\label{fig8}$\phi$ {\it vs} $r$ for $m=0$ at $t=-0.0055$ (solid)$, \, -0.0030$ (dashed) and $-0.0017$ (dotted)}
\end{figure}

\begin{figure}
\includegraphics{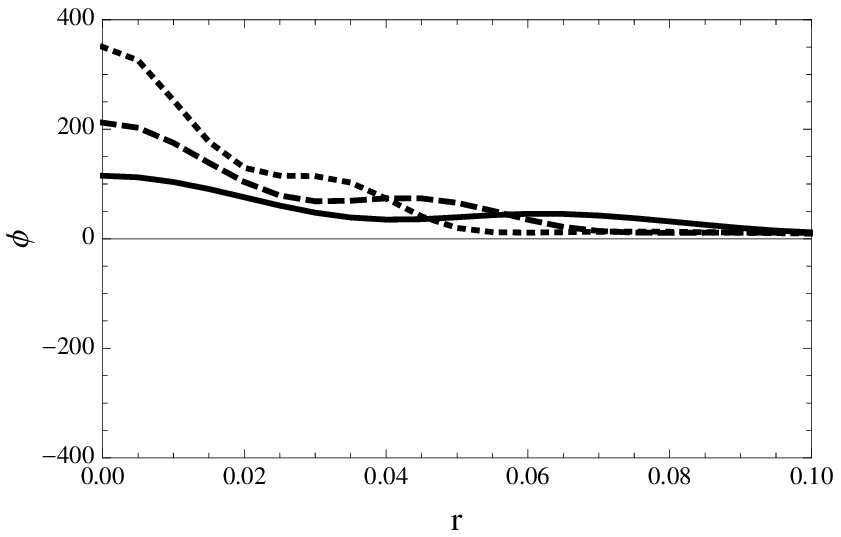}
\caption{\label{fig9}$\phi$ {\it vs} $r$ for $m=1$ at $t=-0.0055$ (solid)$, \, -0.0030$ (dashed) and $-0.0017$ (dotted)}
\end{figure}

\begin{figure}
\includegraphics{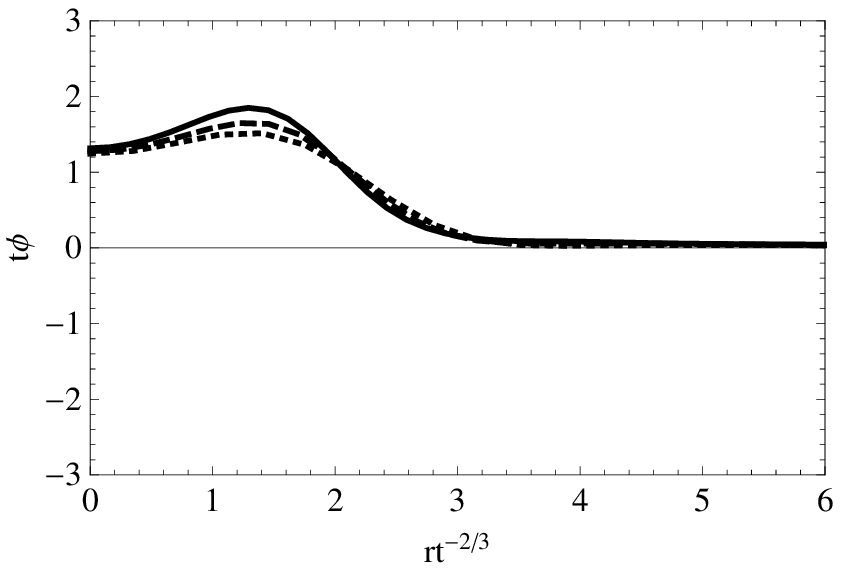}
\caption{\label{fig10}$t\phi$ {\it vs} ${t^{-2/3}}r$ for $m=0$ at 
$t=-0.0055$ (solid)$, \, -0.0030$ (dashed) and $-0.0017$ (dotted)}
\end{figure}

\begin{figure}
\includegraphics{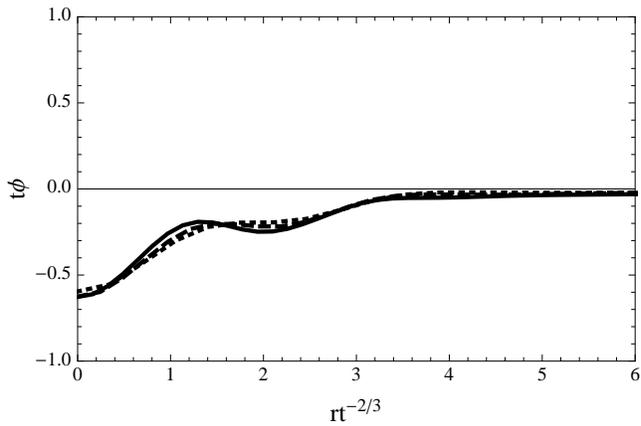}
\caption{\label{fig11}$t\phi$ {\it vs} ${t^{-2/3}}r$ for $m=1$ at 
$t=-0.0055$ (solid)$, \, -0.0030$ (dashed) and $-0.0017$ (dotted)}
\end{figure}

In the simulations, let $N$ be the number of gridpoints and $r_{\rm max}$ be the maximum value
of $r$.  Then we have $\Delta r = {r_{\rm max}}/(N-1)$. 
$N=800$ was used for the runs resulting in figures \ref{fig1}-\ref{fig3}. 
$N=16000$ was used for the runs resulting in figures \ref{fig4}-\ref{fig11}. 
We choose initial data for the scalar field at a time
$t_0$ to take the form $P=0$ and 
\begin{equation}
\phi = -2 {\frac {r-{r_0}} {\sigma ^2}} \exp \left ( {\frac {- {{(r-{r_0})}^2}} {\sigma ^2}} \right )
\end{equation}      
with $S$ given by the derivative of this expression.  Here, $r_0$ and $\sigma$ are constants. 
For all the results displayed, the parameters are 
${t_0} = -20, \, {r_{\rm max}} = 80, \, {r_0}=35, \, \sigma =1.5$, and $M=5$ where $M$ is the
mass of the Schwarzschild black hole. 
This initial configuration of the scalar field is shown in figure \ref{fig1}.
We run the simulations to a time near the singularity.   Figures \ref{fig2} and \ref{fig3} show the 
configuration of the scalar field near this final time.  Both figures are for $t=-0.23$ with
$m=0$ for figure \ref{fig2} and $m=1$ for figure \ref{fig3}.  The initial scalar field can be thought of as a linear combination of a left mover and a right mover.  The right mover escapes from the black hole, while the left mover plunges towards it.  Some part of the left mover scatters off of the black hole, while another part is captured and approaches the singularity.

We are particularly interested in finding
the behavior of the scalar field as the singularity is approached.  In our coordinate system, the singularity is approached as $t \to 0$ but only in the FRW portion of the spacetime.  To examine
the scalar field behavior near the singularity, we plot the scalar field at three
different times near $t=0$ and deduce a trend from those plots.  The scalar field $\phi$  
is plotted 
as a function of $r$ in a small region near $r=0$
for the times $-0.02, \, -0.01$ and $-0.005$ for the $m=0$ case in figure
\ref{fig4} and for the $m=1$ case in figure \ref{fig5}.  It is clear that the trend is that as the
singularity is approached, the scalar field gets a larger amplitude and a steeper profile.  To see
whether this behavior can be understood quantitatively, we turn to the BKL conjecture: the claim that as a generic singularity is approached in a comoving coordinate system, the field equations will be dominated by terms containing derivatives with respect to time, and all other terms can be neglected. To apply this conjecture, we consider the Klein-Gordon equation 
(\ref{kleingordon}) in the untransformed FRW metric and coordinates given by eqns. (\ref{frw}) and (\ref{adust}).
Keeping only terms containing time derivatives, we find
\begin{equation}
{{\tilde t}^{-2}}{\partial _{\tilde t}} ({{\tilde t}^2} {\partial _{\tilde t}} \phi ) = 0.
\end{equation}      
The general solution of this equation is
\begin{equation}
\phi = {{\tilde t}^{-1}} A({\tilde r}) + B ({\tilde r})
\end{equation}

for some functions $A({\tilde r})$ and $B({\tilde r})$.  In particular, the BKL conjecture leads us to expect that as the singularity is approached, ${\tilde t}\phi$ should approach a function of 
$\tilde r$.  Note, that expressed in terms of the $t,r$ coordinates of our simulation, we have
${\tilde t} =t$ and ${\tilde r} = {K^{-1}} {t^{-2/3}} r$.  Thus, we are led to expect that as
the singularity is approached, if we plot $t\phi$ {\it vs} ${t^{-2/3}}r$ the plots should approach some
limiting profile as $t\to 0$.  To see whether that expectation is realized, such plots are given in 
figure \ref{fig6} for $m=0$ and \ref{fig7} for $m=1$.  These figures are consistent with $t\phi$ 
approaching a limit as the singularity is approached.  However, in order to be more definite, we 
need to approach the singularity more closely. 
The results of simulations which show the scalar field at times closer to the singularity are shown in figures
\ref{fig8}-\ref{fig11}.  Here figure \ref{fig8} shows $\phi$ {\it vs} $r$ for
$m=0$ and the times 
$-0.0055, \, -0.0030,$ and $-0.0017$.  Figure \ref{fig9} shows $\phi$ {\it vs} $r$
at the same times for the 
$m=1$ case.  The corresponding rescaled quantities, that is $t\phi$ {\it vs} 
${t^{-2/3}}r$ are shown in figure \ref{fig10} for $m=0$ and in figure \ref{fig11} 
for $m=1$.     

It is clear from figures \ref{fig10} and \ref{fig11} that $t\phi$ approaches a limit as the
singularity is approached.  Since this is just what one would expect from the BKL conjecture, our simulations lend support to the notion that the BKL conjecture holds, not only in the 
extensively studied case of closed cosmologies, but also in the more physically interesting case of asymptotically flat spacetimes.  To explore this issue further, it would be helpful to go beyond the
case of a test scalar field on a background spacetime.  In particular, one could treat the collapse of a self-gravitating scalar field using the method of\cite{racz} and see whether the singularity produced in the collapse satisfies the BKL conjecture.

\ack
The work of DG was supported by NSF grant PHY-0855532 to Oakland University.  
RS and RA are supported by a grant from the US Department of Energy.

\end{document}